\documentclass{aa}
\usepackage{natbib}
\bibpunct{(}{)}{;}{a}{}{,} 
\usepackage{xcolor}
\usepackage{graphicx}
\usepackage{txfonts}
\usepackage{subcaption}         
\usepackage{lscape}
\usepackage{placeins}
\usepackage{hyperref}

\begin{document}

   \title{A comparative study of solar flux emergence and eruptivity in simulations of horizontal versus toroidal magnetic fields}


   \author{V. Karantanis\inst{1}\corrauth{e.karantanis@uoi.gr}
        \and J. Zhuleku\inst{1}\email{j.zhuleku@uoi.gr}
        \and V. Archontis\inst{1}\email{archontis@uoi.gr}
        \and K. Moraitis\inst{1}\email{k.moraitis@uoi.gr}
        }

   \institute{Department of Physics, University of Ioannina, 45110 Ioannina, Greece
   }

   \date{Received April 23, 2026}

  \abstract
   {Magnetic flux emergence is a fundamental driver of eruptive activity in the solar atmosphere. While many numerical studies employed idealized horizontal flux tubes, toroidal tubes provide a more realistic geometry for finite emerging loops with anchored footpoints.}
   {We compare the evolution and eruptive capability of horizontal and toroidal flux tubes under identical initial parameters.} 
   {We performed 3D resistive magnetohydrodynamic (MHD) simulations of the emerging magnetic flux structures to evaluate their respective dynamics}
   {Although the toroidal tube emerges later than in the horizontal case, it produces a higher frequency of eruption-driven jets (four versus two) because the supply of coronal axial flux is sustained. In contrast, the horizontal tube injects magnetic flux and energy more impulsively, driving stronger but less persistent activity and then rapidly stagnating when its atmospheric axial-flux reservoir is depleted. Free magnetic energy builds up after emergence and is released in discrete drops associated with eruptions. The toroidal case exhibits a quasi-cyclic buildup and release pattern, whereas the horizontal case relaxes to a lower-activity state after its early eruptions. The temporal evolution of relative magnetic helicity mirrors the free-energy evolution. Helicity increases with the stressing and twisting of the coronal field during emergence, peaks near eruptive episodes, and decreases as eruptions remove twisted flux, with the toroidal tube maintaining a more persistent helicity budget that supports recurrent events.}
   {Initial flux-tube geometry strongly controls the coronal flux budget and the storage and release of free energy and helicity, and therefore, the frequency and longevity of eruptive phenomena in emergence-driven active regions.}


   \titlerunning{Flux emergence in horizontal vs. toroidal flux tubes.}
   \authorrunning{Karantanis et al.}

   \maketitle
   \nolinenumbers

\section{Introduction}\label{sec:intro}
The solar atmosphere hosts a wide range of dynamic eruptive phenomena, including jets, flares, and coronal mass ejections (CMEs). These events predominantly occur in active regions (ARs), where the magnetic field strength is significantly higher than in the quiet Sun. It is widely accepted that these phenomena arise from the interaction between magnetic flux emerging from the convection zone and the preexisting ambient magnetic field in the corona.

The magnetic field must first traverse from the convection zone to the photosphere. A possible mechanism for this process is the magnetic buoyancy instability \citep{parker1955formation}. Numerical simulations use buoyancy as the primary trigger for magnetic flux emergence \citep{fan2001emergence,archontis2004emergence}. In these simulations, initial conditions typically involve a twisted flux tube with a density deficit to initiate the instability. The initial twist is crucial for maintaining the coherence of the rising tube \citep{syntelis2019successful}. The model geometry ranges from simple infinite horizontal flux tubes \citep{emonet1998physics},\citep{archontis2005three} to toroidal configurations \citep{hood2009emergence}. Both geometries rise and expand to form an $\Omega$-loop, creating an active region where footpoints correspond to opposite-polarity sunspots connected by a polarity-inversion line (PIL). However, the footpoints in the horizontal flux tube model drift apart indefinitely, and this scenario is inconsistent with observations \citep{liu2006magnetic}. Conversely, toroidal geometry anchors the footpoints deep within the convection zone, limiting their separation to a realistic maximum \citep{mactaggart2009emergence}.

Following emergence through the photosphere, the flux tube compresses the surrounding coronal plasma. This interaction initiates eruptive phenomena. Solar jets, first observed by \citet{shibata1993observations}, are attributed to magnetic reconnection between the emerging flux and the oppositely directed ambient coronal field \citep{1977ApJ...216..123H}. This contact forms a current sheet, leading to a topological reconfiguration of the magnetic field and the subsequent release of magnetic energy. This energy is converted into kinetic energy, accelerating plasma to high velocities and thermal energy via Joule dissipation. These dynamics were first investigated in 2D numerical experiments by \citet{yokoyama1996numerical} and later extended to 3D by \citet{fan2001emergence}. A distinct class of ejection is the blowout jet, which is characterized by the expulsion of cooler plasma from the jet's base arch \citep{moore2010dichotomy}. While the precise onset mechanism remains under investigation, the breakout model provides a significant theoretical framework \citep{antiochos1999model}. In this scenario, the separation of flux tube footpoints facilitates the current sheet formation. The eventual ejection of the flux rope results from magnetic reconnection at this current sheet.

We investigate the differences between a horizontal and a toroidal flux tube, specifically regarding the magnetic flux that they can transport in the solar atmosphere, along with their ability to sustain recurrent eruptions. We use a set up similar to that of \citet{archontis2004emergence} for the horizontal and similar to that of \citet{zhuleku2025recurrent} for the toroidal flux tube. We performed two different 3D MHD simulations, one for each flux tube, while also maintaining the same initial conditions for the two experiments. In Section \ref{sec:model-setup} we present a more detailed description of the numerical setup, in Section \ref{sec:results} we present the results of the experiments, and in Section \ref{sec:conclusions} we conclude with a direct comparison between them.

\section{Model setup} \label{sec:model-setup}

We used the Lare3D code of \citet{arber2001staggered} to numerically solve the 3D time-dependent resistive MHD equations in Cartesian geometry for the horizontal and toroidal tubes. The following MHD equations are in dimensionless form, and the basic quantities are density $ \rho $, velocity $ \boldsymbol{u} $, the magnetic field $ \boldsymbol{B} $, and specific energy density $ \epsilon$. Pressure $p $ was derived from the equation of state and current density $ \mathbf{j} $ from Amp\`ere's law $ \boldsymbol{j} = \mathbf{\nabla} \times \boldsymbol{B} $. Gravity had a constant value of $ g = 274 \, \mathrm{m} \, \mathrm{s}^{-2}$ and resistivity $\eta $ also had a uniform value of $\eta = 0.01$ in dimensionless units. We also included the effects of viscous heating $ S_{\mathrm{visc}} $ and Joule dissipation $ Q_{\mathrm{joule}} $, 

\begin{equation}\label{eq:1}
	\frac{\partial \rho}{\partial t} = \nabla \cdot (\rho \boldsymbol{u})
\end{equation}
\begin{equation}
	\rho \ \bigg(\frac{\partial \boldsymbol{u}}{\partial t} + ( \boldsymbol{u} \cdot \nabla ) \boldsymbol{u} \bigg) = - \nabla p + \boldsymbol{j} \times \boldsymbol{B} + \rho g + S_{\mathrm{visc}}
\end{equation}
\begin{equation}
	\rho \ \bigg(\frac{\partial \epsilon}{\partial t} + ( \boldsymbol{u} \cdot \nabla ) \epsilon \bigg) = - p \mathbf{\nabla} \cdot \boldsymbol{u} + \eta \boldsymbol{j}^{2} + Q_{\mathrm{joule}}
\end{equation}
\begin{equation}
	\frac{\partial \boldsymbol{B}}{\partial t} = \mathbf{\nabla} \times ( \boldsymbol{u} \times \boldsymbol{B} ) - \mathbf{\nabla} \times (\eta  \mathbf{\nabla} \times \boldsymbol{B})
\end{equation}
\begin{equation}\label{eq:2}
	p = \rho \, (\gamma - 1) \, \epsilon.
\end{equation}

\begin{figure}[htp!]
    \centering
    \includegraphics[width=0.8\linewidth]{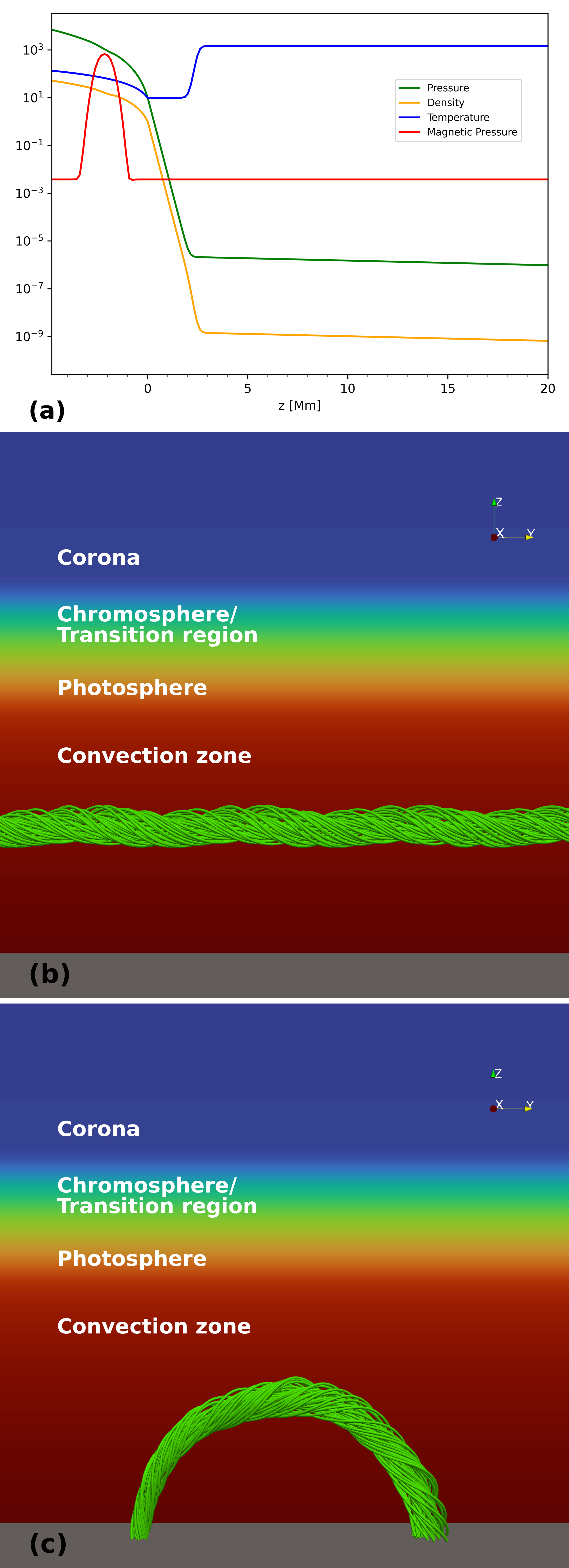}
    \caption{Panel (a): Initial distribution of temperature, pressure and density on the background atmosphere in our simulation in dimensionless units. Panel (b): 3D representation of the twisted magnetic field lines of the horizontal flux tube as is it was placed in the convection zone of the computational box at the beginning of the simulation. Panel (c): 3D representation of the toroidal flux tube as it was initially placed in the convection zone of the computational box at the beginning of the simulation. The gradient color in the background corresponds to the initial density stratification of the environment.}
    \label{fig:initial}
\end{figure}

The normalization we used is based on the values of density, length, and magnetic field at the photosphere. We therefore had $\rho = 1.67 \times 10^{-4} \, \mathrm{kg}\,\mathrm{m}^{-3}$, $H_\mathrm{p} = 0.18 $ Mm the pressure scale height and $B_{\mathrm{ph}} = 300 \, \mathrm{G} $. From these, we calculated the units for pressure $ p = 713.2 \, \mathrm{Pa} $, temperature $T_{0} = 650 \, \mathrm{K} $ , velocity $u_{0} = 2.1 \, \mathrm{km} \,\mathrm{s}^{-1} $ , and the time of each step of the simulation at $t_{0} = 86.9 \, \mathrm{s}$. The initial distribution of the temperature, pressure and density are shown in Figure(\ref{fig:initial}a). We assumed a fully ionized plasma with a specific heat of $\gamma = 5/3 $. 

The code solves equations ~\eqref{eq:1}-~\eqref{eq:2} numerically in their Lagrangian form and then remaps them back onto an Eulerian grid. In both cases, the computational domain had 420 $\times$ 420 $\times$ 420 grid points in each direction, which corresponds to a physical space of 64.6 $\times$ 64.6 $\times$ 64.6 Mm for the $x$, $y$, and $z$ axes.

We assumed periodic boundary conditions in the $y$-direction and open boundary conditions in the $x$-direction and at the top of the computational box. The bottom boundary was set to be closed. The stratification consisted of an adiabatically stratified convection zone at -4.8 $ \leq z < $ 0 Mm, the isothermal photosphere at 0 $ \leq z < $ 1.8 Mm, the chromosphere and transition region at 1.8 $ \leq z < $ 3.2 Mm, and the isothermal corona at 3.2 $ \leq z < $ 59.8 Mm. We also set an ambient coronal field at $B = 10$ G pointing upward while forming an angle of $\theta = 11^{\circ}$ with the $z$-axis and $\phi = 183^{\circ}$ with the $y$-axis. This ensured that the ambient magnetic field lines and the emerging field lines were antiparallel, which triggers reconnection when they converge \citep{galsgaard2007effect}. The same setup of the stratification was used in both flux tubes.

\subsection{Horizontal flux tube}
The geometry of the horizontal flux tube was similar to that of \citet{fan2001emergence}, whose distribution of the magnetic field was given by
\begin{equation}
    \boldsymbol{B} = B_{0} \mathrm{e}^{-r^{2}/R^{2}} \hat{\mathbf{y}} + \alpha r B_{0} \mathrm{e}^{-r^{2}/R^{2}} \hat{\boldsymbol{\theta}},
\end{equation}
where $\hat{\mathbf{y}}$ is the direction of the tube axis, and $\hat{\boldsymbol{\theta}}$ is the azimuthal direction in the tube cross-section. The variable $\alpha$ is the twist parameter of the flux tube, which we set to $\alpha = 0.4$. It represents the number of rotations of the field lines around the axis per unit length. $r = \sqrt{x^{2} + z^{2}}$ is the radial distance from the tube axis, and $R$ is the tube radius, which we set to $R=0.45$ Mm. The magnetic field strength was $B_{0}$ = 6300 G. A 3D representation of the magnetic field of the flux tube in its initial state is shown in Figure (\ref{fig:initial}b). 
The center of the tube was at $z=-2.1$ Mm below the photosphere. Since the flux tube initially was at pressure equilibrium with the environment, we imposed a small density deficit along its axis at the center of the flux tube to destabilize it. 
The horizontal tube was made buoyant by setting the internal temperature of the tube equal to the external. This created a pressure deficit $p_{\mathrm{def}}$ inside the tube to balance the radial component of the Lorentz force and the plasma pressure gradients. The resulting density deficit was then calculated as
\begin{equation}
    \rho_{\mathrm{def}}=\frac{p_{\mathrm{def}}}{T_{z}}=\frac{B^2_{0}\mathrm{e}^{-2r^2/r^2_0}({\alpha}^2 r^2_0-2-2{\alpha}^2 r^2)}{4T_{z}}\mathrm{e}^{-y^2/{\lambda}^2}.
    \label{eq:def}
\end{equation}
We used the parameter $\lambda = 0.9$ Mm, so that only a small part of the tube was buoyant, and we created an emergence of an $\mathrm{\Omega}$-loop \citep{hood2009emergence,mactaggart2009emergence}.

\subsection{Toroidal flux tube}
The model of the toroidal flux tube that we used was similar to that of \citet{hood2009emergence} and had the same initial conditions as the experiment described by \citet{zhuleku2025recurrent}. The Cartesian representation of the components of the initial magnetic field are given as
\begin{equation}
	B_{x} = B_{\theta}(r) \frac{R - R_{0}}{r}
\end{equation}
\begin{equation}
	B_{y} = - B_{\phi}(r) \frac{z - z_{0}}{R} - B_{\theta}(r) \frac{x}{r} \frac{y}{R}
\end{equation}
\begin{equation}
	B_{z} = B_{\phi}(r) \frac{y}{R} - B_{\theta}(r) \frac{x}{r} \frac{z - z_{0}}{R},
\end{equation}
where $R $ is the major radius of the toroidal tube, set to $R_{0} = 2.7$ Mm, and $z_{0}$ the value at the bottom of the computational box, at $z_{0} =$ - 4.8 Mm. The terms $B_{\phi}$ and $B_{\theta}$ are defined for the toroidal as $B_{\phi} = B_{0} e^{-r^{2}/ r_{0}^{2}}$ and $B_{\theta} = \alpha r B_{0} \mathrm{e}^{-r^{2}/r_{0}^{2}}$, with $r_{0} = 0.45$ Mm being the minor radius of the toroidal tube. The toroidal tube was placed in the convection zone at $z=-2.1$ Mm below the photosphere, just like the horizontal tube. The initial distribution of the magnetic field is shown in Figure (\ref{fig:initial}c). 
The toroidal tube was made buoyant in the same way as the horizontal tube, by imposing the temperature to be equal inside and outside of the flux tube. The density deficit in the toroidal tube case was calculated as
\begin{equation}
    \rho_{\mathrm{def}}=\frac{p_{\mathrm{def}}}{T_{z}}=\frac{B^2_{0}\mathrm{e}^{-2 r^2/ r^2_0}({\alpha}^2 r^2_0-2-2 {\alpha}^2 r^2)}{4 T_{z}}.
\end{equation}
This equation is very similar to equation \ref{eq:def} without the last term $\mathrm{e}^{-y^2/ {\lambda}^2}$. Since the parameters such as the initial magnetic field $B_{0}$, twist $a$, and radius $r_{0}$ of the two flux tubes were the same, the initial density deficit was also the same in both cases.

\section{Results \label{sec:results}}

\subsection{Evolution of the emergence}
\begin{figure}[htp!]
    \centering
    \includegraphics[width=\linewidth]{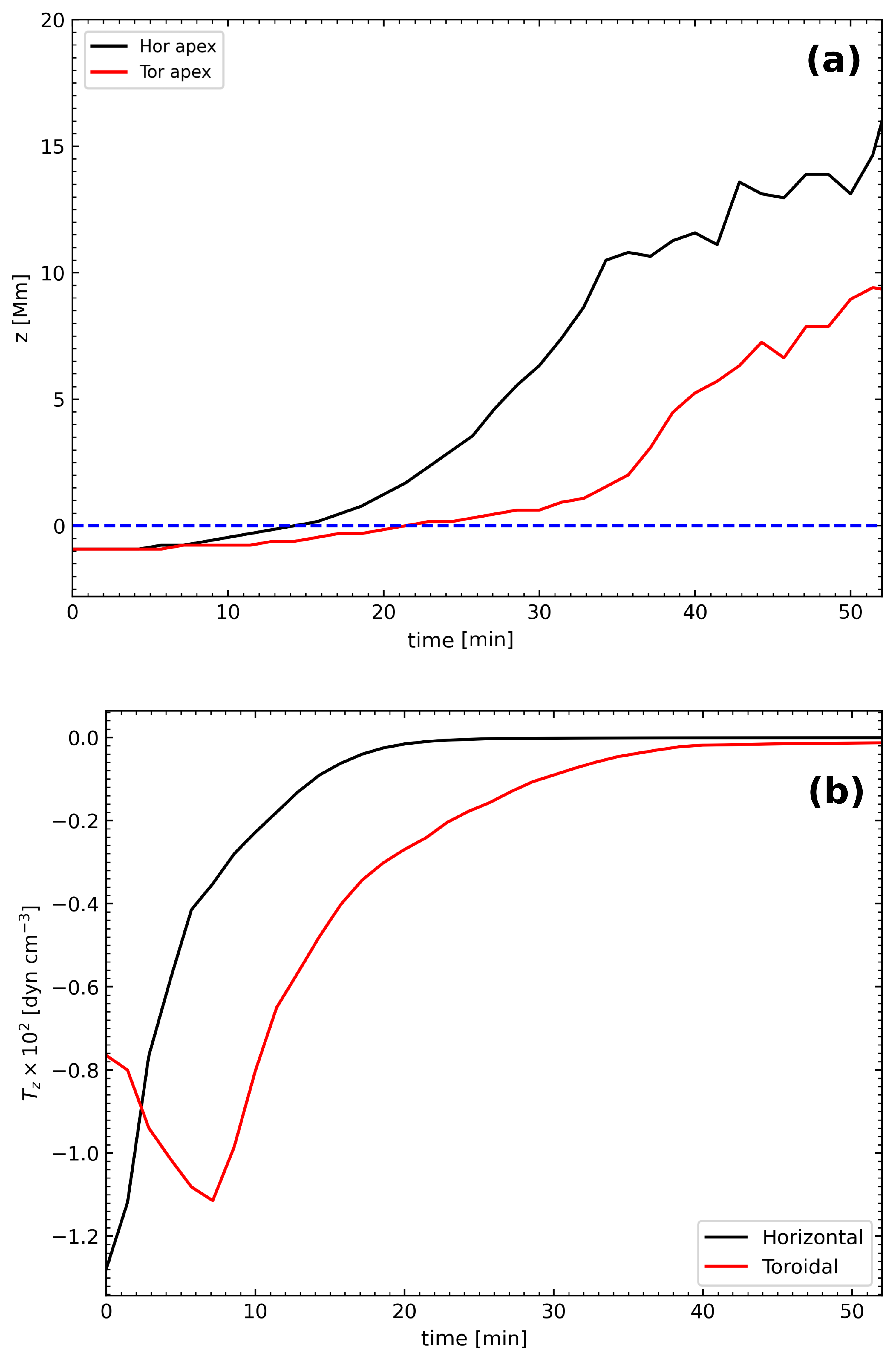}
    \caption{Panel (a): Height-time profile of the flux tube apex. Panel (b): Evolution of the maximum tension in the $ z $ direction. Both panels focus on the beginning of the simulation for the horizontal (black lines) and toroidal flux tube (red lines).}
    \label{fig:height-time-tension}
\end{figure}

To evaluate the dynamics of the emerging plasma, we compare the height-time evolution of the apex for the two flux tube geometries in Figure(\ref{fig:height-time-tension}a). We defined the apex as the upper boundary of the tube where the horizontal magnetic field component, $B_{x}$, dropped below $0.001 \times B_{0}$. A notable difference is observed in their respective emergence rates: the horizontal flux tube reached the photosphere significantly faster than its toroidal counterpart. To analyze this we calculated the $z$-component of the magnetic tension force as,
\begin{equation}
    T_{z} = \frac{1}{4 \pi} ( B_{x} \partial_{x} + B_{y} \partial_{y} + B_{z} \partial_{z} ) \ B_{z} .
\end{equation}
We show the temporal evolution of the maximum tension within the flux tubes in Figure (\ref{fig:height-time-tension}b). Below the tube axis, the tension force is directed upward ($T_{z} \ge 0$), while above the axis, it acts downward ($T_{z} < 0$). Initially, the downward component is dominant, providing a significant restoring force that opposes the buoyant rise. However, as the magnetic pressure inside the tube leads the emergence and the tube undergoes magnetic expansion, the curvature of the upper field lines decreases, subsequently reducing the magnitude of the negative tension component. As the maximum tension approaches zero, the magnetic buoyancy drives the tube upward more effectively. Our results show that the magnetic tension in the horizontal tube decreases much faster than in the toroidal tube, leading to an earlier arrival at the photosphere. This is a direct result of the initial magnetic geometries. The toroidal configuration is inherently strongly curved and has more tightly curved field lines, resulting in a stronger tension force that more persistently resists the emergence than in the horizontal case.

\begin{figure*}[htp!]
    \centering
    \includegraphics[width=\linewidth]{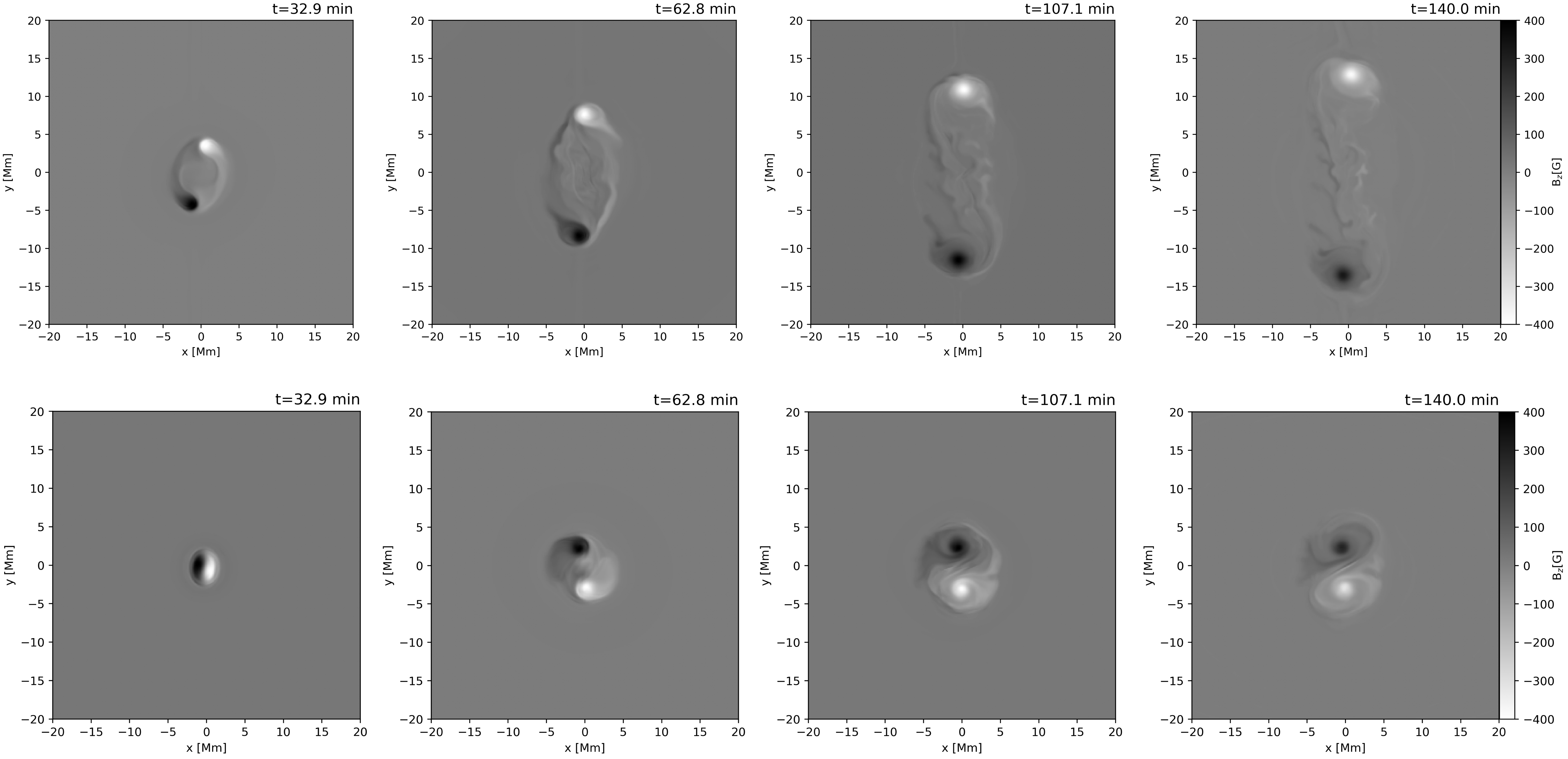}
    \caption{Upper row: Evolution of the vertical component of the magnetic field $B_{z}$ viewed in the $x-y$ plane in the photosphere ($z=0$) for the horizontal tube. Bottom row: Evolution of $B_{z}$ viewed in the $x-y$ plane in the photosphere ($z=0$) for the toroidal tube.}
    \label{fig:Bz_top}
\end{figure*}

As flux emerged in both simulations, it created a characteristic bipolar region at the photospheric level, as shown in Figure (\ref{fig:Bz_top}). To better understand the dynamics of this emergence, we provide a series of snapshots that track the movement of the magnetic footpoints over time. A clear distinction emerges between the two geometries: in the horizontal case, the footpoints continuously drift apart and maintain this divergence until the end of the simulation. The toroidal tube, however, behaves differently. While its footpoints initially separate at a similar rate as the horizontal tube, they eventually reach a maximum separation distance and remain fixed for the rest of the evolution \citep{hood2009emergence},\citep{zhuleku2025recurrent}.
This is due to the initial conditions of the model. The footpoints of the toroidal tube are physically anchored at the bottom boundary, which naturally limits how far the emerging bipolar spots can migrate. In contrast, the horizontal tube was modeled as an infinitely long structure parallel to the surface. Therefore, its footpoints are free to continue their lateral expansion as the magnetic arch grows and fills the atmosphere.

\begin{figure*}[htp!]
    \centering
    \includegraphics[width=\linewidth]{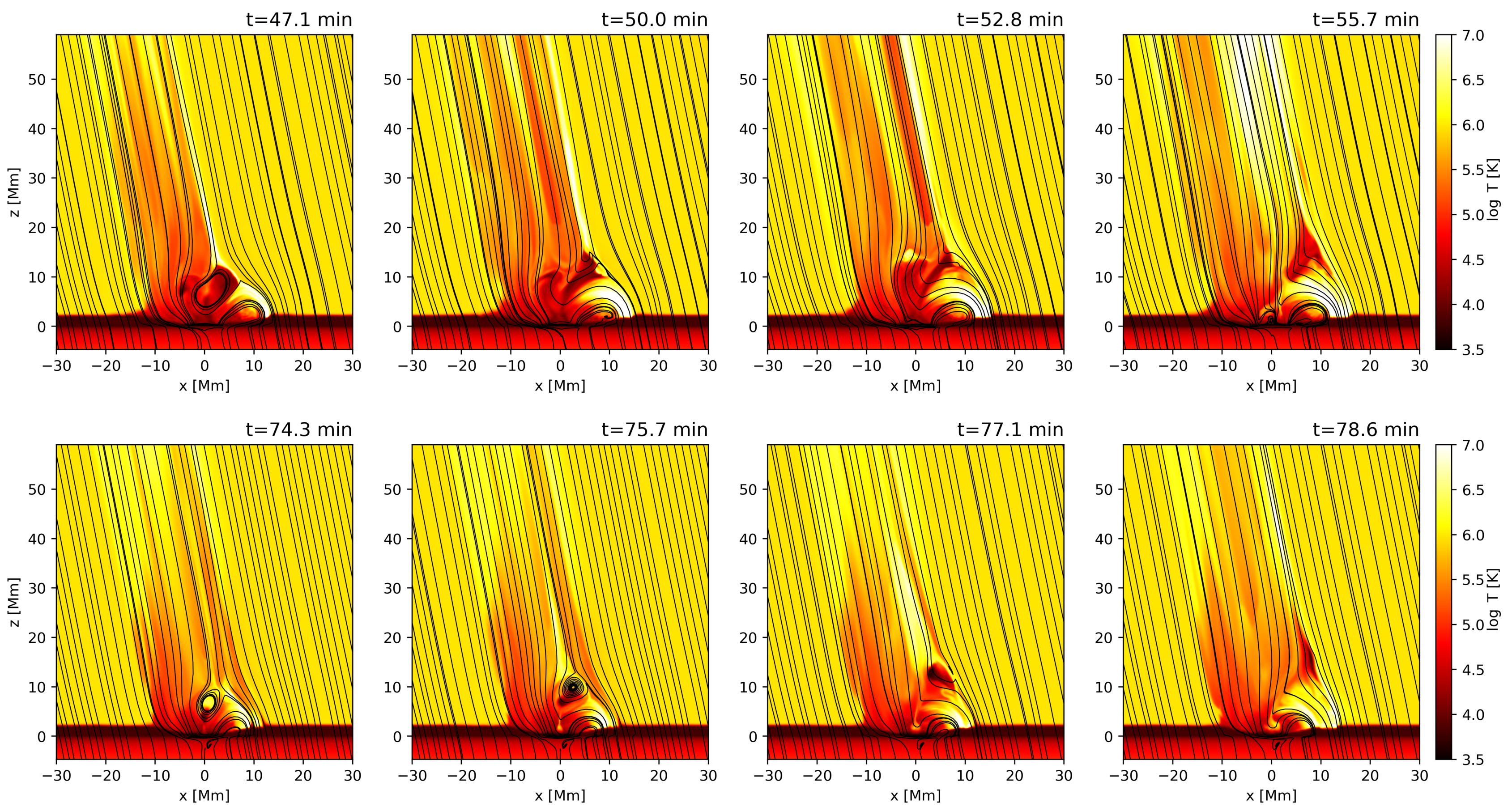}
    \caption{Upper row: Logarithm of temperature of the horizontal flux tube viewed from the side in the $x-z$ midplane along with the magnetic field lines. Bottom row: Logarithm of the temperature of the toroidal flux tube viewed from the side in the $x-z$ midplane along with the magnetic field lines.}
    \label{fig:Temperature}
\end{figure*}

When the flux tube emerges above the photosphere, it adiabatically expands into the isothermal corona. In our setup, the ambient coronal magnetic field is oriented antiparallel to the emerging flux. As the two magnetic systems approach and interact, a current sheet forms at their interface, triggering magnetic reconnection. The resulting high-velocity outflows accelerate plasma to form the inverted-Y jets (Standard jets)\citep{1977ApJ...216..123H, forbes1984numerical}. Both simulations successfully produce these standard jets. The horizontal tube forms its jet at approximately $t=17$ min, whereas the toroidal tube reaches this stage at $t=28$ min. In the horizontal and toroidal cases, coherent flux rope structures form. These flux ropes are generated through internal reconnection between the sheared field lines within the core of the emerging tube \citep{manchester2004eruption,archontis2008eruption},\citep{mactaggart2014magnetic}. When they are formed, these structures can becomes unstable and expelled upward. The eruptions of the flux ropes produce energetic phenomena that are frequently observed in the Sun. When the erupting flux ropes interact with an external ambient magnetic field, the resulting events are known as blowout jets and transport a significant fraction of plasma from the emerging structure into the upper corona. These explosive events eject substantial amounts of magnetic flux and account for a large portion of the total energy released during the emergence process. Figure (\ref{fig:Temperature}) illustrates the atmospheric temperature distribution before and during these eruptive phases. Our simulations show that the frequency of these eruptions depends on the initial geometry of the tubes. We found only two eruptions for the horizontal tube, whereas the toroidal tube produced four. The underlying cause for this difference in eruption frequency constitutes a separate research topic and will be addressed in forthcoming work.

\subsection{Kinetic energy and mass flux}
\begin{figure}[htp!]
    \centering
    \includegraphics[width=\linewidth]{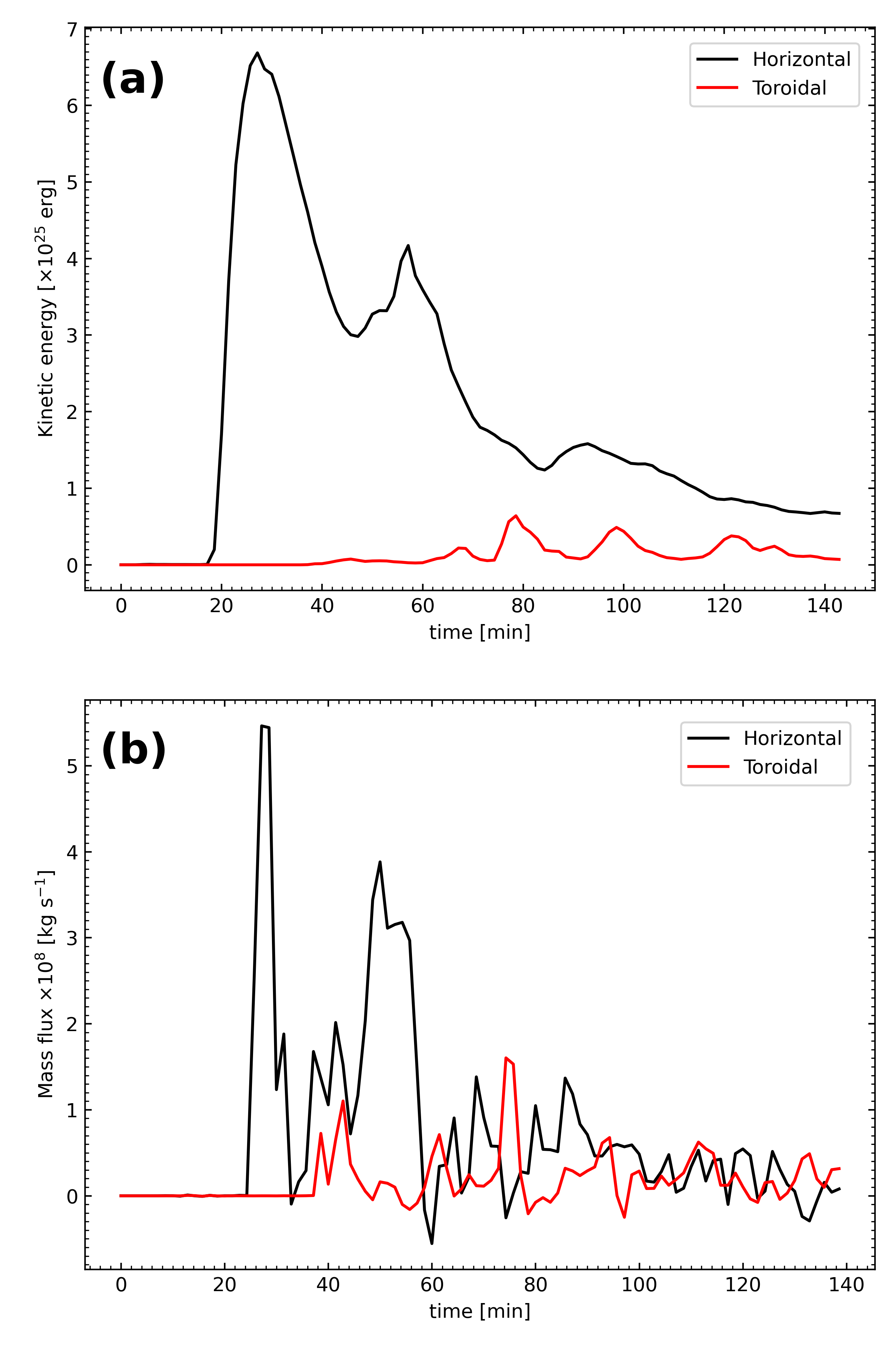}
    \caption{Panel (a): Evolution of the kinetic energy calculated in the volume above $z = 3 \, \mathrm{Mm}$ for the horizontal (black line) and toroidal flux tube (red line). Panel (b): Evolution of the mass flux that passes through the surface at $z = 3.7 \, \mathrm{Mm}$ for the horizontal (black line) and toroidal flux tube (red line).}
    \label{fig:kinetic-mass}
\end{figure}

To quantify the kinetic energy that is released during these eruptive events, we calculated the volume-integrated kinetic energy within the coronal domain ($z=3$ to $z=60$ Mm) using the following expression:
\begin{equation}
E_{\mathrm{kin}} = \iiint \frac{1}{2} \rho u^{2} \ \mathrm{d}x \mathrm{d}y \mathrm{d}z.
\end{equation}
Figure (\ref{fig:kinetic-mass}a) displays the temporal evolution of the total kinetic energy for both simulations. The kinetic energy in the horizontal case is significantly higher, by nearly an order of magnitude, than in the toroidal case. In the horizontal case, the first kinetic energy peak is driven by the high plasma density that increases during the initial emergence phase, combined with the onset of the standard jet. The two following peaks specifically mark the timing of the blowout jets. In contrast, the toroidal tube exhibits much more modest energy signatures. The first increase in kinetic energy, appearing as a subtle bump around $t = 45 \, \mathrm{min} $, coincides with the formation of the standard jet, while the four subsequent peaks correspond to the sequence of blowout eruptions. This suggests that the major curvature and fixed anchoring of the toroidal geometry not only delay emergence, but also act as a physical constraint, limiting the volume of plasma and the total magnetic energy available for conversion into kinetic motion compared to the more expansive horizontal tube.

By integrating the kinetic energy over the duration of the simulation, we found that the horizontal tube released a total of $E_{\mathrm{kin}}=1.74 \times 10^{29}$ $\mathrm{erg}$, whereas the toroidal tube produced $E_{\mathrm{kin}}=1.22 \times 10^{28}$ $\mathrm{erg}$. These values are consistent with the observational range of kinetic energies for large-scale solar jets \citep{raouafi2016solar,schmieder2022solar}. Although the toroidal tube produced more frequent eruptions, the cumulative kinetic energy of the horizontal tube remained higher by an order of magnitude. This suggests that the individual eruptive events in the horizontal case are substantially more energetic. They overcompensate for their lower frequency and dominate the total energy budget of the emergence process. 

To investigate why the horizontal case is so much more energetic, we analyzed the transport of plasma into the corona. The mass flux, $\Phi_{\mathrm{m}}$, through a horizontal surface $S$ at the base of the corona ($z=3.7$ Mm) is calculated as
\begin{equation}
\Phi_{\mathrm{m}} = \iint_{S} \rho u_{z} \ \mathrm{d}x \mathrm{d}y .
\end{equation}
Figure (\ref{fig:kinetic-mass}b) shows the temporal evolution of the mass flux for both cases. The horizontal flux tube increases significantly at approximately $t=28$ min, which agrees with the time during which the magnetic field emerges and expands. This is followed by a sustained mass flux associated with the standard jet and two distinct pulses representing the blowout jets at $t=55$ min and $t=90$ min. Between these eruptive events, a complex pattern of bidirectional flows forms as plasma drains and ascends along the magnetic fieldlines \citep{innes1997bi}.
In contrast, the flux transport in the toroidal tube is much more limited. The initial magnetic flux emerges as a small peak at $t=40$ min, followed closely by the standard jet at $t=45$ min. While the four subsequent blowout jets (occurring after $t=60$ min) increase the mass flux, the magnitudes remain far below those of the horizontal case.
This result provides a physical explanation for the kinetic energy disparity discussed previously because the horizontal tube is capable of transporting a significantly larger volume of dense plasma into the corona. This increased mass loading, combined with high velocity outflows, results in the order of magnitude difference in total kinetic energy.

\subsection{Magnetic flux}
Active regions are characterized by intense magnetic field concentrations. They are the primary sites for solar flares and more violent eruptions such as coronal mass ejections (CMEs). A defining feature of these events is their dependence on magnetic twist. This twist is a vital precondition for eruption because it stores the free magnetic energy that is necessary to drive reconnection and the subsequent conversion into kinetic and thermal energy \citep{aschwanden2017global}.
One way to quantify the magnetic energy that is transported to the corona is to measure the axial magnetic flux. In our coordinate system, the axial flux represents the $B_{y}$ component passing through a vertical cross-section (a plane perpendicular to the solar surface). Because only a fraction of the subsurface flux actually emerges above the photosphere \citep{cheung2014flux}, calculating the emerged axial flux allows us to determine which of the two magnetic flux tubes is more efficient at transporting magnetic energy into the upper atmosphere.

\begin{figure}[htp!]
     \centering
     \includegraphics[width=\linewidth]{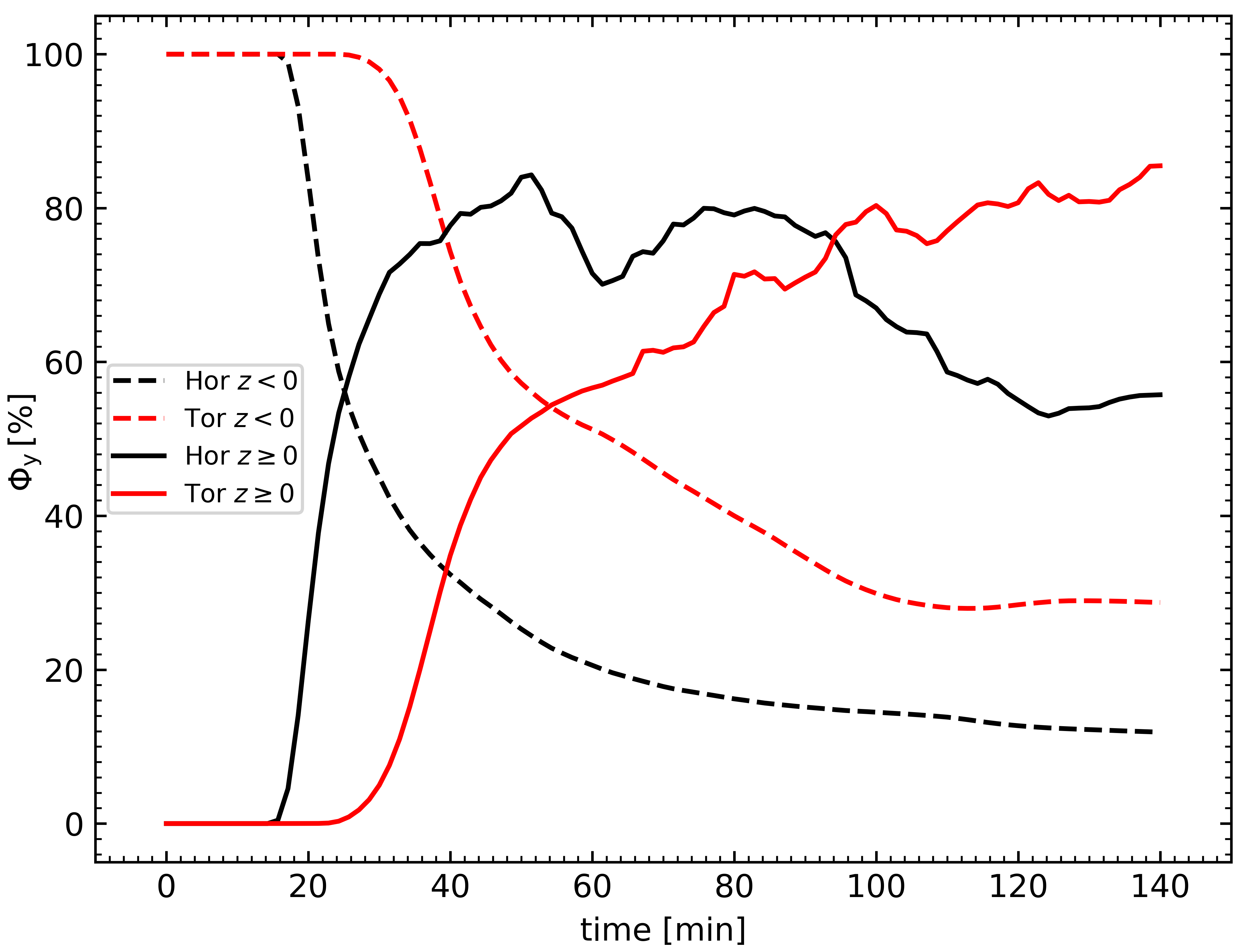}
     \caption{Percentage of the axial flux ($\Phi_{y}$) that passes through the $x-z$ surface for the horizontal and toroidal flux tubes. The total axial flux calculated below the photosphere as a function of time is denoted by $z<0$, and $z \ge 0$ refers to the total axial flux at and above the photosphere.}
     \label{fig:phi_y}
 \end{figure}

We began by examining the evolution of the axial magnetic flux for both cases. To distinguish between the subsurface and emerged components, we show the calculated normalized percentage of the flux integrated over the $\mathrm{x}-\mathrm{z}$ midplane below and above the photosphere in Figure (\ref{fig:phi_y}).
In our results, the axial flux within the solar interior ($z < 0$) is represented by dashed lines, while the flux that has emerged into the photosphere and above ($z \ge 0$) is shown with solid lines.
We calculated the axial flux as
\begin{equation}
    \Phi_{y}^{z<0} = \iint_{z<0} B_{y} \ \mathrm{d}x \ \mathrm{d}z \ , \ \ \Phi_{y}^{z \ge 0} = \iint_{z \ge 0} B_{y} \ \mathrm{d}x \ \mathrm{d}z.
\end{equation}
 
\begin{figure}[htp!]
     \centering
     \includegraphics[width=\linewidth]{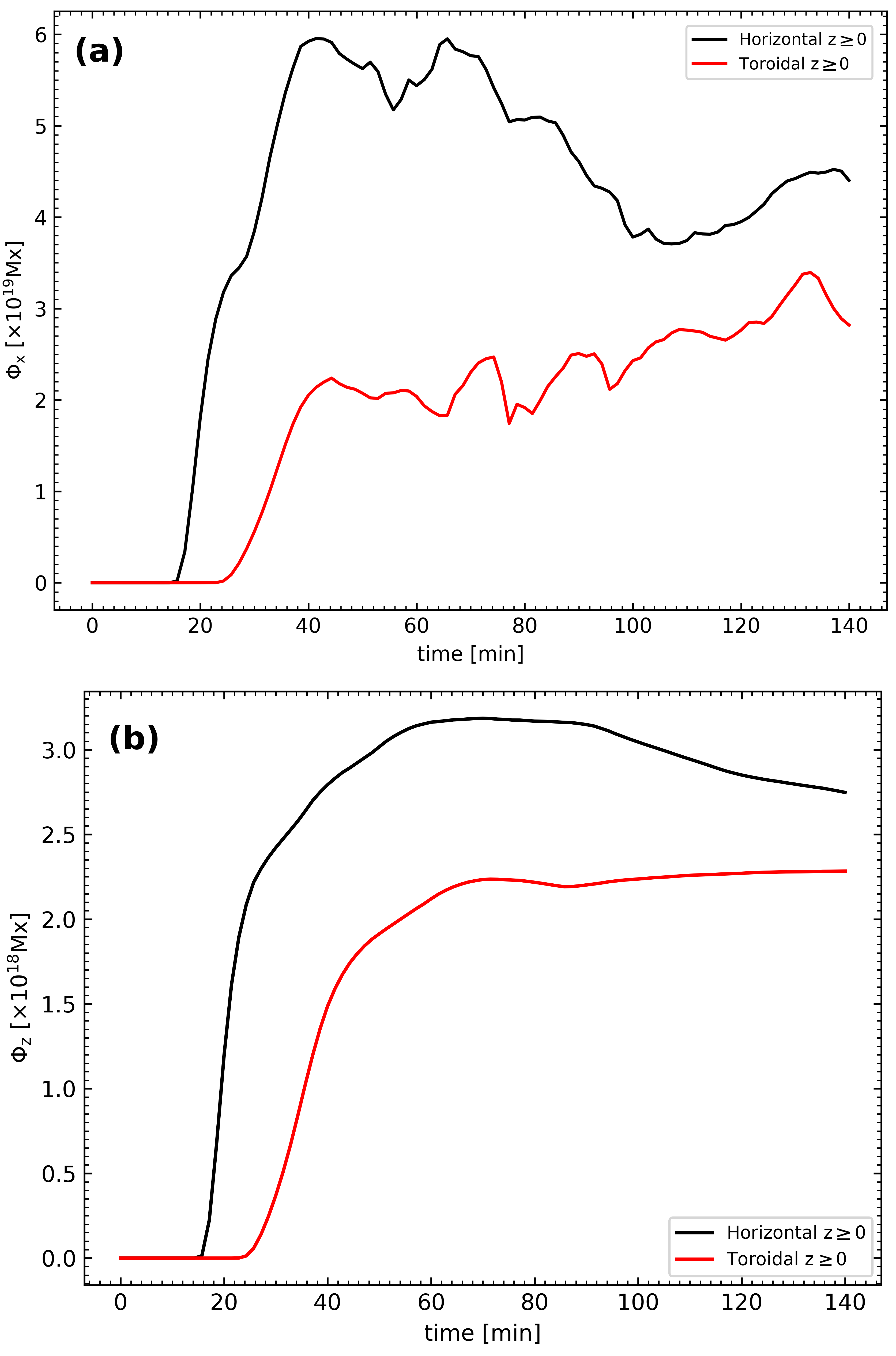}
     \caption{Panel (a): Flux that passes through the $y-z$ surface ($\Phi_{x}$). The horizontal flux tube brings more flux in the $x$-direction than the toroidal. Panel (b): Flux that passes through the $x-y$ surface ($\Phi_{z}$). The horizontal transports more vertical flux than the toroidal. The two surfaces are restricted to heights at and above the photosphere.}
     \label{fig:phi_x_z}
 \end{figure}

A clear dichotomy in emergence dynamics is visible in Figure (\ref{fig:phi_y}). The horizontal flux tube (solid black line) exhibits an impulsive emergence phase as it begins to rise rapidly at $t=15$ min, reaching a maximum value at $t=50$ min. However, this peak is transient since following the eruption of the main blowout jets, the atmospheric flux significantly decays and drops by nearly 30\% by the end of the simulation. This suggests that the flux loss of the horizontal tube is greater, likely through magnetic reconnection or expulsion from the domain \citep{archontis2013numerical,moreno2013plasma}.

In contrast, the toroidal flux tube (solid red line) displays a delayed but more sustained emergence. It begins to rise at $t=30$ min with a shallower slope than in the horizontal case. Interestingly, the axial flux of the toroidal tube does not decay to the same degree, but instead continues to accumulate in the atmosphere, eventually forming a stable plateau. Notably, at $t=95$ min, a crossover occurs, where the atmospheric axial flux of the toroidal tube surpasses that of the horizontal tube. This indicates that although the toroidal case hinders the initial speed of emergence, it ultimately supports a more robust and long-lasting accumulation of magnetic flux in the solar corona.

A significant result is observed in the evolution of the subsurface axial flux in Figure (\ref{fig:phi_y}). For the horizontal case (dashed black line), the magnetic reservoir is nearly completely depleted. The interior flux drops to nearly 10\% of its initial amount by the end of the simulation. This indicates that approximately 90\% of the initial magnetic flux is successfully transported out of the solar interior and into the atmosphere. In the horizontal case, being  buoyant along its entire length, the axial flux escapes almost entirely from the convection zone.

In contrast, in the toroidal tube (dashed red line), the depletion of the axial flux saturates at roughly 30\% of its initial amount, although it initially decreased in response to the emergence of the loop apex. This means that a significant fraction of the toroidal flux remains trapped below the photosphere. The curve flattens out and ceases to drop further after $t=100$ min. This result provides direct quantitative evidence of the fact that the anchored toroidal tube footpoints prevent the full emergence of the axial magnetic flux and permanently keep a large portion of the structure inside the solar interior.

The emergence of the axial magnetic flux into the solar atmosphere is a fundamental prerequisite for the generation of eruptive events. The quantity of this emerged flux directly governs the system capacity to form flux ropes and drive eruptions.

In the case of the horizontal flux tube, the atmospheric axial flux decreases rapidly following each eruptive episode. After the initial surge of the flux emergence concludes, the system lacks a mechanism to replenish the axial flux required to create more flux ropes. Consequently, the horizontal tube exhausts its magnetic energy early, rendering it incapable of producing further blowout jets.

Conversely, the toroidal flux tube retains a substantial reservoir of axial flux even after the primary emergence phase has plateaued. We attribute this to the fundamental difference in transport dynamics since the horizontal tube impulsively dumps the majority of its magnetic energy into the corona, leading to a swift stagnation of activity. 

In contrast, the toroidal tube injects flux at a gradual, sustained rate. This steady supply allows the toroidal system to drive a higher frequency of eruptions, albeit of lower individual intensity, and to maintain a sufficient axial flux budget to fuel recurrent activity long after the horizontal case has become dormant.

To fully capture the three-dimensional expansion of the emerging structures, we also show the magnetic flux passing through the $y-z$ midplane and the horizontal $x-y$ plane in Figure (\ref{fig:phi_x_z}), calculated using the following integrals: 
\begin{equation}
    \Phi_{x} = \iint_{z \ge 0} B_{x} \ \mathrm{d}y \ \mathrm{d}z \ 
\end{equation}
\begin{equation}
    \Phi_{z} = \iint B_{z} \ \mathrm{d}x \ \mathrm{d}y \ \bigg|_{z=0 Mm} - \ \iint B_{z} \ \mathrm{d}x \ \mathrm{d}y \ \bigg|_{z=59.8 Mm}.
\end{equation}
Figure (\ref{fig:phi_x_z}a) illustrates the transverse flux ($\Phi_{x}$) at and above the photosphere. The time evolution shows that the transverse flux in the horizontal tube is higher for almost the entire duration of the numerical experiment. This indicates that the magnetic field perpendicular to its axis in the horizontal tube is much more effectively dispersed and dominates the atmospheric volume early in the emergence process.
This trend is further supported by the vertical flux ($\Phi_{z}$) at the photospheric level, shown in Figure (\ref{fig:phi_x_z}b). The horizontal tube initially facilitates a much higher rate of upward flux transport until both systems eventually stabilize after $t=60$ min. The rapid surge in perpendicular and vertical fluxes for the horizontal case reinforces our previous conclusion that while the toroidal tube erupts more frequently, its late emergence and geometric constraints effectively restrain large volumes of flux, preventing the flux from escaping the main structure as efficiently as in the horizontal tube.
\subsection{Free energy}
\begin{figure}[htp!]
    \centering
     \includegraphics[width=\linewidth]{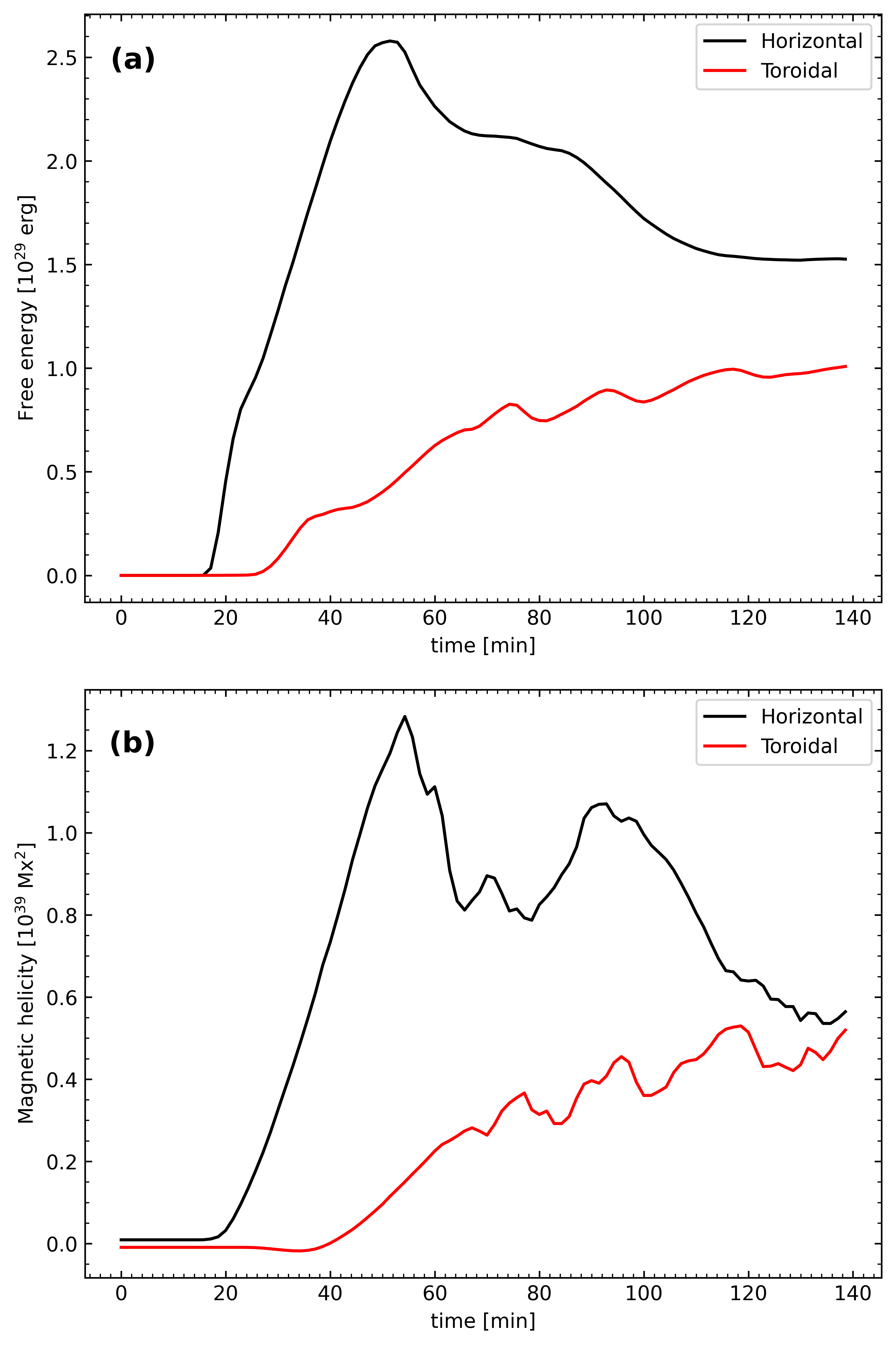}
     \caption{Panel (a): Temporal evolution of the free magnetic energy. Panel (b): Temporal evolution of the relative magnetic helicity. Both panels refer to the coronal volume and show the horizontal (black lines) and toroidal flux tube (red lines).}
     \label{fig:free-energy-helicity}
\end{figure}

Since a requirement for an emerging flux to trigger the eruption is enough available free energy stored in the flux tube configuration, we expect the release of energy from the system at every eruption. We therefore calculated the free energy ($E_{\mathrm{f}}$) in the system as the difference between the total magnetic energy and the energy of the potential magnetic field,
\begin{equation}
    E_{\mathrm{f}} = \frac{1}{{8 \pi}} \int_V ( {\boldsymbol{B}^{2}} \ - \ \boldsymbol{B}_{\mathrm{p}}^{2} ) \ \mathrm{d}V,
\end{equation}
in the volume $V$ above the photosphere. Here, $\boldsymbol{B}_{\mathrm{p}}$ is the potential magnetic field, which was chosen to have the same normal distribution  in the boundary of the volume as in the original magnetic field, and was computed as in \citet{moraitis2014validation}. In Figure (\ref{fig:free-energy-helicity}a) we plot the free energy through time. The free energy of the horizontal flux tube starts to increase abruptly at a time of around $t=20$ minutes, which is when the flux tube emerges into the corona. From this point onward until a time of $t=50$ minutes, free energy builds up in the system. After this point, an eruption expels plasma upward, which reduces the total amount of free energy in the remaining configuration. Another eruption later starts at $t=93$ minutes. This reduces the amount of free energy even further, until it saturates at around $E_{\mathrm{f}} = 1.5 \times 10^{29} $ erg. Since there are neither more eruptions nor a mechanism that substitutes flux, the remaining free energy remains unchanged. 

The free energy in the toroidal flux tube also starts to increase by the time the tube emerges into the corona, but later than in the horizontal flux tube. Nevertheless, when enough energy has build up, the toroidal flux tube erupts and removes a substantial amount of free energy. However, unlike the horizontal tube, free energy increases until another eruption expels it once again. This cycle continues, and we observed a total of four eruptions. At the end of the simulation, the amount of free energy in the system appeared to stabilize at around $E_{\mathrm{f}} \sim 10^{29} $ erg. \citet{zhuleku2025recurrent} used the same setup and let the simulation run for twice as long as we did. They observed two more eruptions before the free energy saturated at a slightly lower value.

\subsection{Magnetic helicity}
In addition to the free energy, magnetic helicity can provide information about eruptions. The degree to which magnetic field lines are twisted and tangle around each other is quantified by the magnetic helicity. For the magnetic fields in the Sun, the relative helicity is calculated following \citet{berger1984topological,finn1985magnetic},
\begin{equation}
H_\mathrm{r}= \int \big( \boldsymbol{A} + \boldsymbol{A}_{\mathrm{p}} \big) \cdot \big( \boldsymbol{B} - \boldsymbol{B}_{\mathrm{p}} \big) \ \mathrm{d}V,
\end{equation}
where $\boldsymbol{A}$ and $\boldsymbol{A}_{\mathrm{p}}$ are the vector potentials of the original and potential magnetic fields, satisfying $\boldsymbol{B}= \nabla \times \boldsymbol{A}$ and $\boldsymbol{B}_{\mathrm{p}}= \nabla \times \boldsymbol{A}_{\mathrm{p}}$, respectively. We plot the relative helicity ($H_\mathrm{r}$) over the whole simulation for the horizontal and toroidal flux tubes in Figure (\ref{fig:free-energy-helicity}b). Just like in the case of magnetic energy, helicity follows the same increase as either flux tube emerges in the corona due to the stress and twist of magnetic field lines \citep{valori2016magnetic}. Peaks in relative helicity coincide with peaks in free energy, which correspond to each eruptive event we observed. As helicity is removed through the upper boundary, the value of relative helicity decreases with each event and is restored afterward, allowing for more eruptions, especially in the case of the toroidal flux tube. Particularly in the case of the horizontal tube, the relative helicity drops significantly after the second eruption, which means that the remaining structure in the corona is less twisted. At around the same time, the free energy also decreases, which might explain why there are no more eruptions. A similar behavior as in the value of relative helicity and free energy is observed for the later stages of the simulation of the toroidal flux tube, as shown by \citet{zhuleku2025recurrent}. We highlight that recent studies by \citet{mactaggart2025bound} and \citet{yeates2026energy} showed that the relative magnetic helicity provides a lower bound for the free energy of the system, and the ratio $E_{\mathrm{f}} / H_\mathrm{r}$ would give more insight into the energy-helicity relation. We aim to explore the effect of this relation on the evolution of our system in a future study.

\section{Conclusions} \label{sec:conclusions}
We presented a controlled comparison of two widely used initial magnetic flux tube geometries for magnetic flux emergence experiments: an initially horizontal buoyant flux tube, and a toroidal flux tube. They both evolved under the same stratification, boundary conditions, resistivity, and tube parameters using Lare3D. The goal was to isolate how the initial tube geometry alone modifies (i) the emergence timescale, (ii) the magnetic flux budget in the atmosphere, and (iii) the frequency and energetics of eruptive episodes (standard and blowout jets).

Our findings are summarized below.
\begin{itemize}
    \item The horizontal tube reaches the photosphere and corona earlier than the toroidal tube. This delay in the toroidal case is consistent with the higher curvature-related magnetic tension, which persists longer and opposes the buoyant rise until magnetic expansion reduces the downward-tension component.
    \item Both configurations form bipolar photospheric regions, but their subsequent evolution diverges: the horizontal-tube polarities continue to drift apart throughout the simulation, whereas the toroidal case reaches a maximum separation and then remains approximately fixed. This behavior directly follows from the anchored geometry of the toroidal tube, which limits lateral expansion and yields a stabler long-lived bipole configuration.
    \item Both simulations produce a standard jet followed by blowout jets when the emerging field reconnects with the antiparallel ambient coronal field and internal reconnection forms flux-rope structures. In our runs, the toroidal tube generated roughly twice as many blowout eruptions as the horizontal tube, indicating that a toroidal geometry more readily supports recurrent eruptive cycle \citep{zhuleku2025recurrent}.
    \item Despite fewer eruptions, the horizontal case attains substantially higher coronal kinetic energy (and higher mass flux through the base of the corona). This indicates that the horizontal tube transports (and accelerates) a larger amount of dense plasma into the corona during its more impulsive emergence phase, producing fewer but more energetic episodes than the toroidal case.
    \item The horizontal tube emerges its subsurface axial-flux reservoir more efficiently and rapidly, but the atmospheric axial flux subsequently declines after major eruptive episodes, consistent with significant flux removal via reconnection/expulsion. In contrast, the toroidal tube shows delayed emergence, but maintains (and continues to build toward) a more sustained atmospheric axial-flux plateau. This behavior is consistent with a slow-release injection of axial flux that can replenish the coronal flux budget between eruptions and thereby support recurrent activity.
    \item The buildup and release of coronal free magnetic energy reflects the distinct eruptive cadence of the two geometries. In the horizontal case, the free energy rises rapidly after the magnetic field enters the corona ($t\approx 20$ min), and it then decreases in two major eruptive episodes and subsequently saturates at $E_{\mathrm{f}}\simeq 1.5\times 10^{29}$~erg when no further eruptions occur. In contrast, the toroidal case shows a delayed but sustained free-energy injection, $E_{\mathrm{f}}$ increases after coronal emergence, is repeatedly reduced by successive eruptions, and continues to rebuild between events, producing a longer-lived cycle of activity before stabilizing at $E_{\mathrm{f}} \sim 10^{29}$~erg.
    \item The temporal evolution of relative magnetic helicity closely tracks the free-energy budget and provides a physically consistent explanation for why the horizontal system becomes inactive earlier. In both configurations, helicity increases as twisted fieldlines are transported into the corona, and the peaks in helicity coincide with peaks in free energy and with the onset of eruptive events. However, after the second eruption in the horizontal case, the relative helicity drops significantly, the post-eruption coronal structure remains significantly less twisted, and the system does not recover sufficient helicity/free energy to trigger additional eruptions. By contrast, the toroidal configuration exhibits a more effective restoration of helicity between eruptions, consistent with its ability to maintain recurrent blowout activity over longer times.

\end{itemize}
Our results suggest that while the horizontal flux tube model is highly effective for simulating singular large-scale explosive events, the toroidal geometry represents the recurrent cycle observed in many solar active regions far more accurately. The trade-off between the impulsive energy release of horizontal tubes and the sustained periodic eruptions of toroidal tubes highlights that the initial geometry of a magnetic structure is a primary determinant of its eruptive behavior in the solar corona.

As an extension of this work, the exact mechanism that causes the toroidal flux tube to generate axial flux and sustain its eruptivity remains to be understood. A parametric study of the magnitude of the magnetic field might also provide further information about the role of the geometry of the two flux tubes regarding their ability to transport magnetic flux in the solar atmosphere.

\vspace{1cm}

\begin{acknowledgements}
      We would like to thank the Referee for their valuable comments. This work has been supported by the European Research Council through the Synergy Grant No.810218 (“The Whole Sun”, ERC-2018-SyG). All the numerical simulations were performed on the ARIS HPC of the National Infrastructures for Research and Technology S.A.(GRNET S.A.) under the project name: Magnetic flux emergence in the Sun and the onset of explosive events.
\end{acknowledgements}

\bibliographystyle{aa} 
\bibliography{sample701}

@ARTICLE{arber2001staggered,
       author = {{Arber}, T.~D. and {Longbottom}, A.~W. and {Gerrard}, C.~L. and {Milne}, A.~M.},
        title = "{A Staggered Grid, Lagrangian-Eulerian Remap Code for 3-D MHD Simulations}",
      journal = {Journal of Computational Physics},
         year = 2001,
        month = jul,
       volume = {171},
       number = {1},
        pages = {151-181},
          doi = {10.1006/jcph.2001.6780},
       adsurl = {https://ui.adsabs.harvard.edu/abs/2001JCoPh.171..151A}
}

@ARTICLE{shibata1993observations,
       author = {{Shibata}, Kazunari and {Ishido}, Yoshinori and {Acton}, Loren W. and {Strong}, Keith T. and {Hirayama}, Tadashi and {Uchida}, Yutaka and {McAllister}, Alan H. and {Matsumoto}, Ryoji and {Tsuneta}, Saku and {Shimizu}, Toshifumi and {Hara}, Hirohisa and {Sakurai}, Takashi and {Ichimoto}, Kiyoshi and {Nishino}, Yohei and {Ogawara}, Yoshiaki},
        title = "{Observations of X-Ray Jets with the Yohkoh Soft X-Ray Telescope}",
      journal = {\pasj},
         year = 1992,
        month = nov,
       volume = {44},
       number = {5},
        pages = {L173-L179},
          doi = {10.1093/pasj/44.5.L173},
       adsurl = {https://ui.adsabs.harvard.edu/abs/1992PASJ...44L.173S}
}

@ARTICLE{1977ApJ...216..123H,
       author = {{Heyvaerts}, J. and {Priest}, E.~R. and {Rust}, D.~M.},
        title = "{An emerging flux model for the solar phenomenon.}",
      journal = {\apj},
         year = 1977,
        month = aug,
       volume = {216},
        pages = {123-137},
          doi = {10.1086/155453},
       adsurl = {https://ui.adsabs.harvard.edu/abs/1977ApJ...216..123H}
}

@ARTICLE{yokoyama1996numerical,
       author = {{Yokoyama}, Takaaki and {Shibata}, Kazunari},
        title = "{Numerical Simulation of Solar Coronal X-Ray Jets Based on the Magnetic Reconnection Model}",
      journal = {\pasj},
         year = 1996,
        month = apr,
       volume = {48},
        pages = {353-376},
          doi = {10.1093/pasj/48.2.353},
       adsurl = {https://ui.adsabs.harvard.edu/abs/1996PASJ...48..353Y}
}

@article{emonet1998physics,
  title={The physics of twisted magnetic tubes rising in a stratified medium: two-dimensional results},
  author={Emonet, T and Moreno-Insertis, F},
  journal={\apj},
  volume={492},
  number={2},
  pages={804--821},
  year={1998}
}

@article{archontis2005three,
  title={The three-dimensional interaction between emerging magnetic flux and a large-scale coronal field: reconnection, current sheets, and jets},
  author={Archontis, Vasilis and Moreno-Insertis, F and Galsgaard, Klaus and Hood, Alan William},
  journal={\apj},
  volume={635},
  number={2},
  pages={1299},
  year={2005},
  publisher={IOP Publishing}
}

@ARTICLE{parker1955formation,
       author = {{Parker}, Eugene N.},
        title = "{The Formation of Sunspots from the Solar Toroidal Field.}",
      journal = {\apj},
         year = 1955,
        month = mar,
       volume = {121},
        pages = {491},
          doi = {10.1086/146010},
       adsurl = {https://ui.adsabs.harvard.edu/abs/1955ApJ...121..491P}
}

@article{fan2001emergence,
  title={The emergence of a twisted $\Omega$-tube into the solar atmosphere},
  author={Fan, Y},
  journal={\apj},
  volume={554},
  number={1},
  pages={L111},
  year={2001},
  publisher={IOP Publishing}
}

@article{archontis2004emergence,
  title={Emergence of magnetic flux from the convection zone into the corona},
  author={Archontis, V and Moreno-Insertis, F and Galsgaard, K and Hood, A and O'shea, E},
  journal={\aap},
  volume={426},
  number={3},
  pages={1047--1063},
  year={2004},
  publisher={EDP Sciences}
}

@article{manchester2004eruption,
  title={Eruption of a buoyantly emerging magnetic flux rope},
  author={Manchester IV, W and Gombosi, T and DeZeeuw, D and Fan, Y},
  journal={\apj},
  volume={610},
  number={1},
  pages={588},
  year={2004},
  publisher={IOP Publishing}
}

@article{archontis2008eruption,
  title={Eruption of magnetic flux ropes during flux emergence},
  author={Archontis, V and T{\"o}r{\"o}k, Tibor},
  journal={\aap},
  volume={492},
  number={2},
  pages={L35--L38},
  year={2008},
  publisher={EDP Sciences}
}

@article{syntelis2019successful,
  title={Successful and failed flux tube emergence in the solar interior},
  author={Syntelis, P and Archontis, V and Hood, A},
  journal={\apj},
  volume={874},
  number={1},
  pages={15},
  year={2019},
  publisher={IOP Publishing}
}

@article{hood2009emergence,
  title={The emergence of toroidal flux tubes from beneath the solar photosphere},
  author={Hood, Alan William and Archontis, Vasilis and Galsgaard, K and Moreno-Insertis, F},
  journal={\aap},
  volume={503},
  number={3},
  pages={999--1011},
  year={2009},
  publisher={EDP Sciences}
}

@article{mactaggart2009emergence,
  title={On the emergence of toroidal flux tubes: general dynamics and comparisons with the cylinder model},
  author={MacTaggart, David and Hood, Alan W},
  journal={\aap},
  volume={507},
  number={2},
  pages={995--1004},
  year={2009},
  publisher={EDP Sciences}
}

@ARTICLE{mactaggart2014magnetic,
       author = {{MacTaggart}, D. and {Haynes}, A.~L.},
        title = "{On magnetic reconnection and flux rope topology in solar flux emergence}",
      journal = {\mnras},
         year = 2014,
        month = feb,
       volume = {438},
       number = {2},
        pages = {1500-1506},
          doi = {10.1093/mnras/stt2285},
archivePrefix = {arXiv},
       eprint = {1311.4225},
 primaryClass = {astro-ph.SR},
       adsurl = {https://ui.adsabs.harvard.edu/abs/2014MNRAS.438.1500M}
}

@article{liu2006magnetic,
  title={The magnetic field, horizontal motion and helicity in a fast emerging flux region which eventually forms a delta spot},
  author={Liu, Jihong and Zhang, Hongqi},
  journal={\solphys},
  volume={234},
  pages={21--40},
  year={2006},
  publisher={Springer}
}

@article{galsgaard2007effect,
  title={The effect of the relative orientation between the coronal field and new emerging flux. I. Global properties},
  author={Galsgaard, K and Archontis, V and Moreno-Insertis, F and Hood, AW},
  journal={\apj},
  volume={666},
  number={1},
  pages={516},
  year={2007},
  publisher={IOP Publishing}
}

@article{forbes1984numerical,
  title={Numerical simulation of reconnection in an emerging magnetic flux region},
  author={Forbes, TG and Priest, ER},
  journal={\solphys},
  volume={94},
  pages={315--340},
  year={1984},
  publisher={Springer}
}

@article{moore2010dichotomy,
  title={Dichotomy of solar coronal jets: standard jets and blowout jets},
  author={Moore, Ronald L and Cirtain, Jonathan W and Sterling, Alphonse C and Falconer, David A},
  journal={\apj},
  volume={720},
  number={1},
  pages={757},
  year={2010},
  publisher={IOP Publishing}
}

@article{antiochos1999model,
  title={A model for solar coronal mass ejections},
  author={Antiochos, SK and DeVore, CR and Klimchuk, JA},
  journal={\apj},
  volume={510},
  number={1},
  pages={485},
  year={1999},
  publisher={IOP Publishing}
}

@article{zhuleku2025recurrent,
	title={Recurrent eruptions from the emergence of a toroidal flux tube},
	author={Zhuleku, J and Archontis, V and Moraitis, K},
	journal={\apj},
	volume={986},
	number={1},
	pages={47},
	year={2025},
	publisher={IOP Publishing}
	}

@article{innes1997bi,
	title={Bi-directional plasma jets produced by magnetic reconnection on the Sun},
	author={Innes, DE and Inhester, B and Axford, WI and Wilhelm, K},
	journal={\nat},
	volume={386},
	number={6627},
	pages={811--813},
	year={1997},
	publisher={Nature Publishing Group UK London}
}

@article{aschwanden2017global,
  title={Global energetics of solar flares. V. Energy closure in flares and coronal mass ejections},
  author={Aschwanden, Markus J and Caspi, Amir and Cohen, Christina MS and Holman, Gordon and Jing, Ju and Kretzschmar, Matthieu and Kontar, Eduard P and McTiernan, James M and Mewaldt, Richard A and O’Flannagain, Aidan and others},
  journal={\apj},
  volume={836},
  number={1},
  pages={17},
  year={2017},
  publisher={The American Astronomical Society}
}

@ARTICLE{cheung2014flux,
       author = {{Cheung}, Mark C.~M. and {Isobe}, Hiroaki},
        title = "{Flux Emergence (Theory)}",
      journal = {Living Reviews in Solar Physics},
         year = 2014,
        month = dec,
       volume = {11},
       number = {1},
          eid = {3},
        pages = {3},
          doi = {10.12942/lrsp-2014-3},
       adsurl = {https://ui.adsabs.harvard.edu/abs/2014LRSP...11....3C}
}

@article{moreno2013plasma,
  title={Plasma jets and eruptions in solar coronal holes: a three-dimensional flux emergence experiment},
  author={Moreno-Insertis, Fernando and Galsgaard, Klaus},
  journal={\apj},
  volume={771},
  number={1},
  pages={20},
  year={2013},
  publisher={IOP Publishing}
}

@ARTICLE{berger1984topological,
       author = {{Berger}, M.~A. and {Field}, G.~B.},
        title = "{The topological properties of magnetic helicity}",
      journal = {Journal of Fluid Mechanics},
         year = 1984,
        month = oct,
       volume = {147},
        pages = {133-148},
          doi = {10.1017/S0022112084002019},
       adsurl = {https://ui.adsabs.harvard.edu/abs/1984JFM...147..133B}
}

@ARTICLE{finn1985magnetic,
       author = {{Finn}, John M. and {Antonsen}, Jr., Thomas M.},
        title = "{Magnetic helicity: What is it and what is it good for?}",
      journal = {Comments on Plasma Physics and Controlled Fusion},
         year = 1985,
        month = may,
       volume = {9},
        pages = {111-126},
       adsurl = {https://ui.adsabs.harvard.edu/abs/1985CoPPC...9..111F}
}

@article{valori2016magnetic,
  title={Magnetic helicity estimations in models and observations of the solar magnetic field. Part I: Finite volume methods},
  author={Valori, Gherardo and Pariat, Etienne and Anfinogentov, Sergey and Chen, Feng and Georgoulis, Manolis K and Guo, Yang and Liu, Yang and Moraitis, Kostas and Thalmann, Julia K and Yang, Shangbin},
  journal={\ssr.},
  volume={201},
  pages={147--200},
  year={2016},
  publisher={Springer}
}

@article{archontis2013numerical,
	title={A numerical model of standard to blowout jets},
	author={Archontis, Vasilis and Hood, AW},
	journal={\apjl},
	volume={769},
	number={2},
	pages={L21},
	year={2013},
	publisher={IOP Publishing}
}

@ARTICLE{raouafi2016solar,
       author = {{Raouafi}, N.~E. and {Patsourakos}, S. and {Pariat}, E. and {Young}, P.~R. and {Sterling}, A.~C. and {Savcheva}, A. and {Shimojo}, M. and {Moreno-Insertis}, F. and {DeVore}, C.~R. and {Archontis}, V. and {T{\"o}r{\"o}k}, T. and {Mason}, H. and {Curdt}, W. and {Meyer}, K. and {Dalmasse}, K. and {Matsui}, Y.},
        title = "{Solar Coronal Jets: Observations, Theory, and Modeling}",
      journal = {\ssr},
         year = 2016,
        month = nov,
       volume = {201},
       number = {1-4},
        pages = {1-53},
          doi = {10.1007/s11214-016-0260-5},
archivePrefix = {arXiv},
       eprint = {1607.02108},
 primaryClass = {astro-ph.SR},
       adsurl = {https://ui.adsabs.harvard.edu/abs/2016SSRv..201....1R}
}

@ARTICLE{schmieder2022solar,
       author = {{Schmieder}, Brigitte},
        title = "{Solar Jets: SDO and IRIS Observations in the Perspective of New MHD Simulations}",
      journal = {Frontiers in Astronomy and Space Sciences},
         year = 2022,
        month = feb,
       volume = {9},
          eid = {820183},
        pages = {820183},
          doi = {10.3389/fspas.2022.820183},
archivePrefix = {arXiv},
       eprint = {2201.11541},
 primaryClass = {astro-ph.SR},
       adsurl = {https://ui.adsabs.harvard.edu/abs/2022FrASS...920183S}
}

@ARTICLE{mactaggart2025bound,
       author = {{MacTaggart}, David},
        title = "{A bound for relative magnetic helicity in terms of free magnetic energy}",
      journal = {Discover Space},
         year = 2025,
        month = dec,
       volume = {129},
       number = {1},
          eid = {19},
        pages = {19},
          doi = {10.1007/s11038-025-09578-8},
       adsurl = {https://ui.adsabs.harvard.edu/abs/2025DiSpa.129...19M}
}

@article{yeates2026energy,
  title={Energy Bounds from Relative Magnetic Helicity in Spherical Shells},
  author={Yeates, Anthony R and Hornig, Gunnar},
  journal={\apj},
  volume={999},
  number={1},
  pages={46},
  year={2026},
  publisher={The American Astronomical Society}
}

@article{moraitis2014validation,
  title={Validation and benchmarking of a practical free magnetic energy and relative magnetic helicity budget calculation in solar magnetic structures},
  author={Moraitis, K and Tziotziou, K and Georgoulis, MK and Archontis, V},
  journal={\solphys},
  volume={289},
  number={12},
  pages={4453--4480},
  year={2014},
  publisher={Springer}
}

\end{document}